\documentclass[twocolumn,superscriptaddress,nofootinbib,notitlepage,longbibliography,aps]{revtex4-1}
\pdfoutput=1

\usepackage{graphicx}
\usepackage{amsmath}
\usepackage{amssymb}
\usepackage{amsfonts}
\usepackage[dvipsnames]{xcolor}
\usepackage[linktoc=none]{hyperref}
\usepackage[utf8]{inputenc}

\hypersetup{colorlinks,linkcolor={blue},citecolor={teal},urlcolor={violet}}

\newcommand{\documenttitle}{Direct Detection of Light Dark Matter from Evaporating Primordial Black Holes}

\newcommand{\changed}[1]{{\color{black}#1}}

\newcommand{\INFN}{INFN - Sezione di Napoli, Complesso Univ. Monte S. Angelo, I-80126 Napoli, Italy}
\newcommand{\UNINA}{Dipartimento di Fisica ``Ettore Pancini'', Università degli studi di Napoli ``Federico II'', Complesso Univ. Monte S. Angelo, I-80126 Napoli, Italy}
\newcommand{\SSM}{Scuola Superiore Meridionale, Università degli studi di Napoli ``Federico II'', Largo San Marcellino 10, 80138 Napoli, Italy}

\begin{document}

\title{\documenttitle}

\author{Roberta Calabrese}
\affiliation{\UNINA}
\affiliation{\INFN}
\author{Marco Chianese}
\email{chianese@na.infn.it}
\affiliation{\UNINA}
\affiliation{\INFN}
\author{Damiano F.G. Fiorillo}
\affiliation{\UNINA}
\affiliation{\INFN}
\author{Ninetta Saviano}
\affiliation{\INFN}
\affiliation{\SSM}

\date{\today}
\begin{abstract}
    The direct detection of sub-GeV dark matter interacting with nucleons is hampered by the low recoil energies induced by scatterings in the detectors. This experimental difficulty is avoided in the scenario of boosted dark matter where a component of dark matter particles is endowed with large kinetic energies. In this Letter, we point out that the current evaporation of primordial black holes with masses from $10^{14}$ to $10^{16}$ g is a source of boosted light dark matter with energies of tens to hundreds of MeV. Focusing on the XENON1T experiment, we show that these relativistic dark matter particles could give rise to a signal orders of magnitude larger than the present upper bounds. Therefore, we are able to significantly constrain the combined parameter space of primordial black holes and sub-GeV dark matter. In the presence of primordial black holes with a mass of $10^{15}~\mathrm{g}$ and an abundance compatible with present bounds, the limits on DM-nucleon cross-section are improved by four orders of magnitude.
\end{abstract}

\maketitle


{\bf Introduction.---} 
Though Dark Matter (DM) is one of the backbones of the standard cosmological model, its nature is still unknown~\cite{Bertone:2018krk}. So far, all the evidences of its existence are related only to its gravitational imprints, while we have no clue about its non-gravitational interactions~\cite{Kahlhoefer:2017dnp,PerezdelosHeros:2020qyt}. Among the different ways to probe DM properties, direct detection experiments are achieving very stringent constraints on DM-nucleon~\cite{Angloher_2005, Aprile_2012, DarkSide:2014llq, Angloher:2016ooq, DARWIN:2016hyl,SuperCDMS:2016wui, DarkSide-20k:2017zyg,XENON:2018voc,LUX-ZEPLIN:2018poe,XENON:2019gfn, XENON:2019zpr,CRESST:2017ues,CRESST:2019jnq, XENON:2020kmp,COSINE-100:2021xqn,PandaX:2021osp} and DM-electron~\cite{Essig:2011nj, Essig:2012yx, Essig:2017kqs, Crisler:2018gci, DarkSide:2018ppu, DAMIC:2019dcn, Andersson:2020uwc} interactions (see Ref.~\cite{Schumann:2019eaa} for a recent review). These experiments search for the nuclear and electron recoil energy caused by the possible scatterings with DM particles that surround us. Due to a rapidly decreasing sensitivity at low recoil energies, the constraints on DM interactions dramatically weaken for DM masses smaller than about 1 GeV (1 MeV) for DM interactions with nuclei (electrons), thus leaving light DM candidates poorly explored by direct searches. There are complementary approaches to the exploration of light DM. Fermionic DM particles lighter than 0.1~keV are highly constrained by phase-space arguments~\cite{Boyarsky:2008ju,DiPaolo:2017geq} and structure formation~\cite{Tremaine:1979we}. Other model-dependent constraints are placed by astrophysical and cosmological observations, as well as by colliders~\cite{Cohen:2015toa,Daci:2015hca,Ali-Haimoud:2015pwa,Baryakhtar:2017dbj,Bondarenko:2019vrb,Coogan:2021sjs,Gluscevic:2017ywp,Xu:2018efh,Slatyer:2018aqg,Nadler:2019zrb}. 

The experimental limitation of direct detection can be circumvented in the framework of boosted dark matter, where a fraction of DM particles gains a velocity higher that the virial one due to a number of different mechanisms~\cite{Agashe:2014yua,Giudice:2017zke,Fornal:2020npv}. In recent years, it has been pointed out that light DM particles can be upscattered to (semi-)relativistic velocities through collisions with cosmic-rays~\cite{Cappiello:2018hsu,Bringmann:2018cvk,Ema:2018bih,Cappiello:2019qsw,Ema:2020ulo,PROSPECT:2021awi}. The unavoidable presence of such a subdominant boosted DM component has improved the model-independent constraints on the DM-nucleon cross-section at the level of $\sim 10^{-31}~{\rm cm^2}$.

In this Letter, we propose the evaporation of Primordial Black Holes (PBHs) at present times as a source of boosted light DM particles (hereafter denoted as $\chi$). Throughout this work, we refer to this scenario as ePBH-DM. PBHs are hypothetical black holes formed soon after the inflationary epoch through the gravitational collapse of density fluctuations in the early universe~\cite{Zeldovich:1967lct, Harrison:1969fb}. As a result of combining quantum field theory and general relativity, PBHs emit Hawking radiation in the form of graybody spectrum with a temperature $T_{\rm PBH}$ inversely proportional to the PBH mass~\cite{hawking1974black,hawking1975particle, zeldovich1976charge, Carr:1976zz, Page:1976df, Page:1977um, PhysRevD.41.3052}. PBHs with a mass $M_\mathrm{PBH}=\mathcal{O}(10^{15}~\mathrm{g})$ are evaporating at the present times, process that has been employed to set stringent upper bounds on the PBHs abundance~\cite{Boudaud:2018hqb,Laha:2019ssq,Ballesteros:2019exr,Dasgupta:2019cae,Coogan:2020tuf,Calabrese:2021zfq,DeRomeri:2021xgy,Baker:2021btk} (see Ref.~\cite{Carr:2020gox} for a comprehensive review). DM particles with a mass $m_\chi< T_\mathrm{PBH}$ are also emitted with (semi)-relativistic momenta by PBHs. This mechanism has been vastly studied as a way to produce DM particles in the early Universe~\cite{Morrison:2018xla,Baldes:2020nuv,Bernal:2020kse,Bernal:2020ili,Bernal:2020bjf,Auffinger:2020afu,Gondolo:2020uqv,Masina:2021zpu,Cheek:2021odj,Cheek:2021cfe}. However, the production of DM particles in the present Universe, to our knowledge, has never been investigated.

\begin{figure}[t!]
    \centering
    \includegraphics[width=\columnwidth]{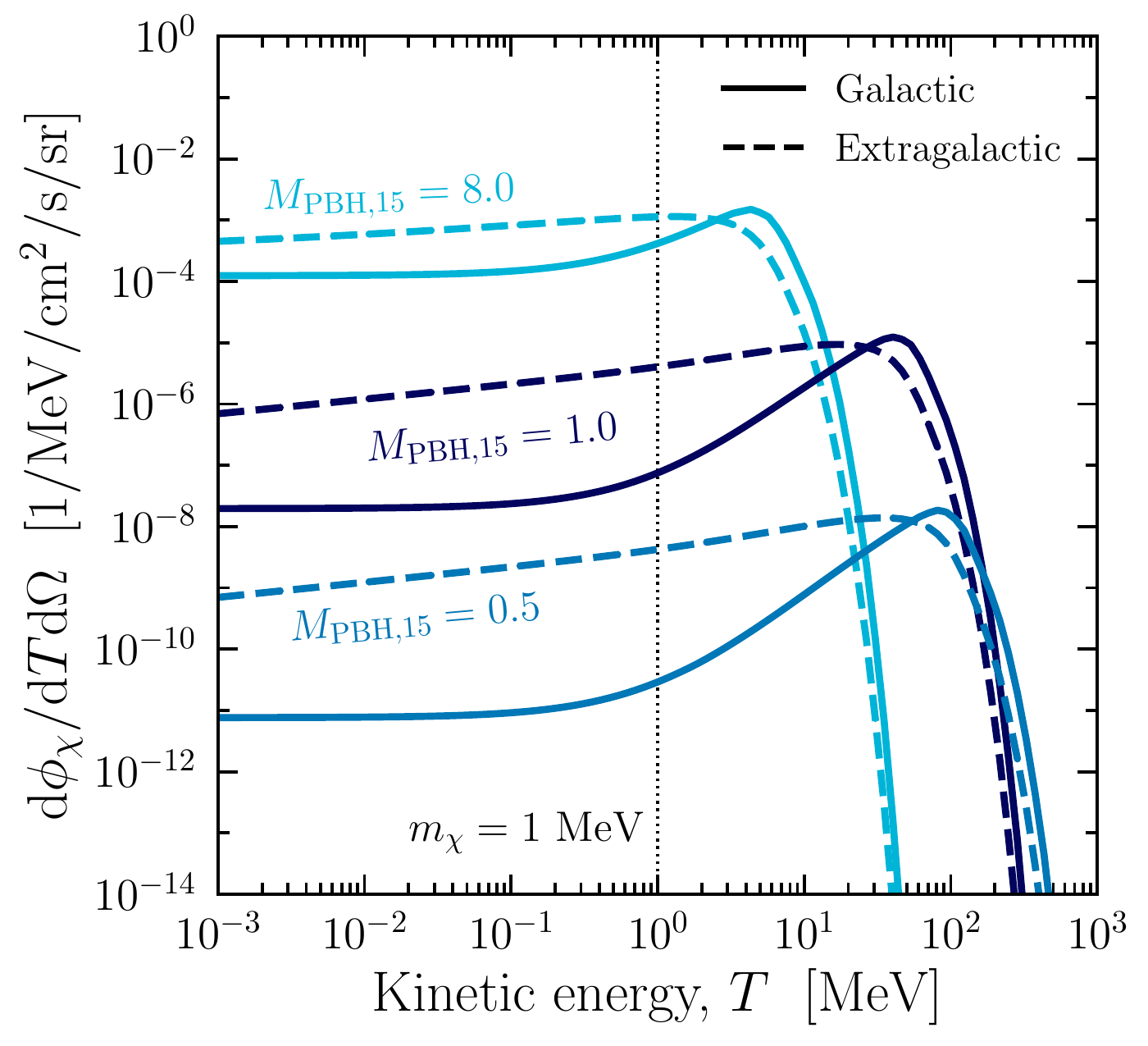}
    \caption{{\bf Diffuse DM flux from evaporating PBHs.} We show the diffuse flux of DM particles incident at the Earth as a function of the kinetic energy. The galactic (solid) and extragalactic (dashed) lines are shown separately. Different colors are used to identify different PBH masses ($M_{\text{PBH},15}=M_{\text{PBH}}/10^{15}$ g). The PBH abundance is chosen as the maximum value allowed by the present constraints~\cite{Carr:2020gox}: $f_{\text{PBH}}$ is equal to $2.9 \times 10^{-10}$, $3.9 \times 10^{-7}$, $3.7 \times 10^{-4}$ for a PBH mass of $0.5$, $1.0$, and $8.0$ in units of $10^{15}~{\rm g}$, respectively. Effects of energy losses in Earth and atmosphere are not included.}
    \label{fig:flux}
\end{figure}

\begin{figure}[t!]
    \centering
    \includegraphics[width=\columnwidth]{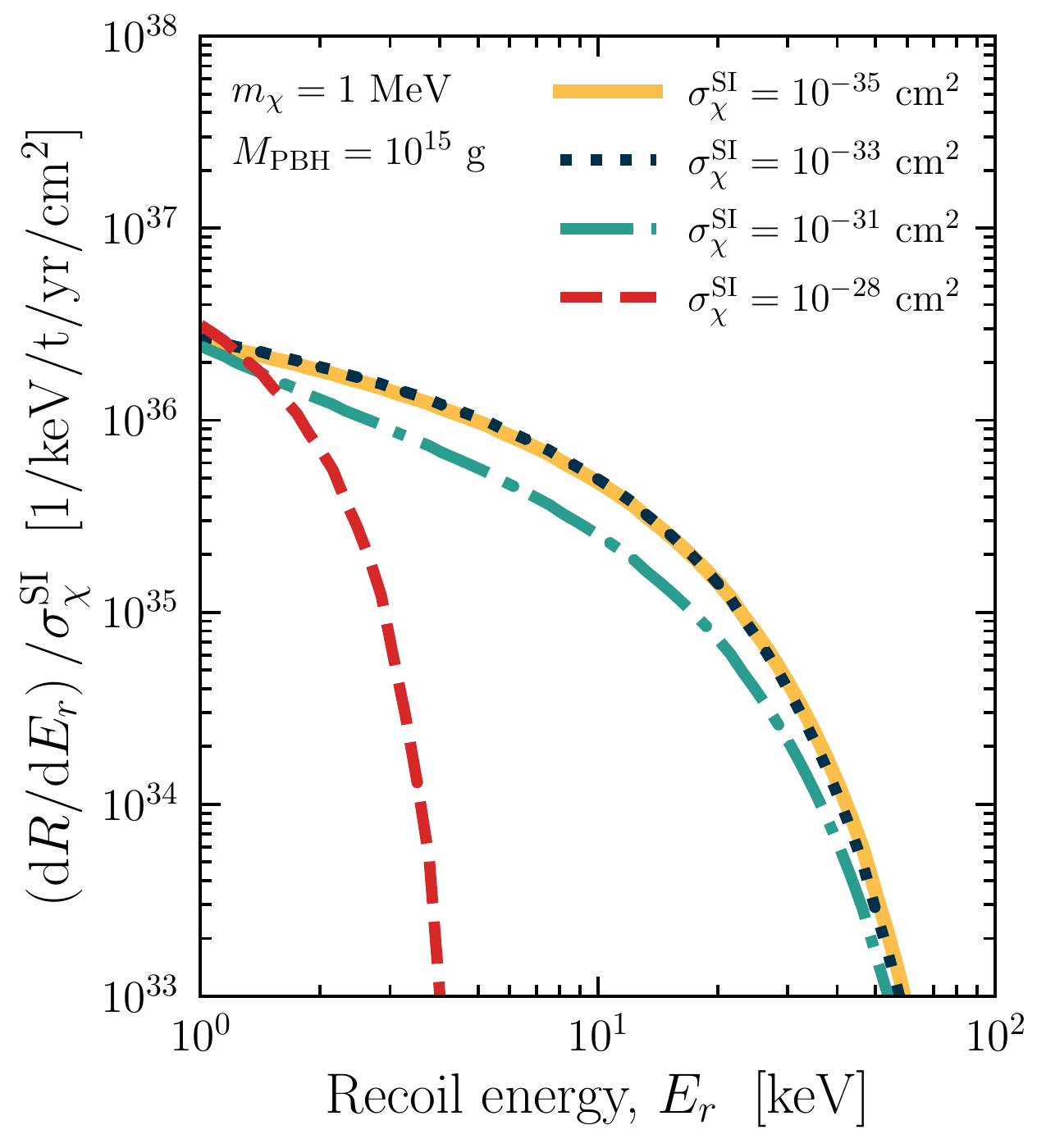}
    \caption{{\bf Ratio of XENON1T event rate over DM-nucleon cross-section.} We show the ratio of the event rate induced by DM scatterings inside the XENON1T detector over the DM-nucleon cross-section. Here, we account for the effects of propagation through the Earth and the atmosphere. The lines correspond to different values of the DM-nucleon cross-section. The remaining parameters are chosen as $M_{\text{PBH}}=10^{15}$~g, $f_{\text{PBH}}=3.9\times10^{-7}$, and $m_\chi=1$~MeV.}
    \label{fig:rate}
\end{figure}

\begin{figure*}[tbh!]
    \centering
    \includegraphics[width=0.485\textwidth]{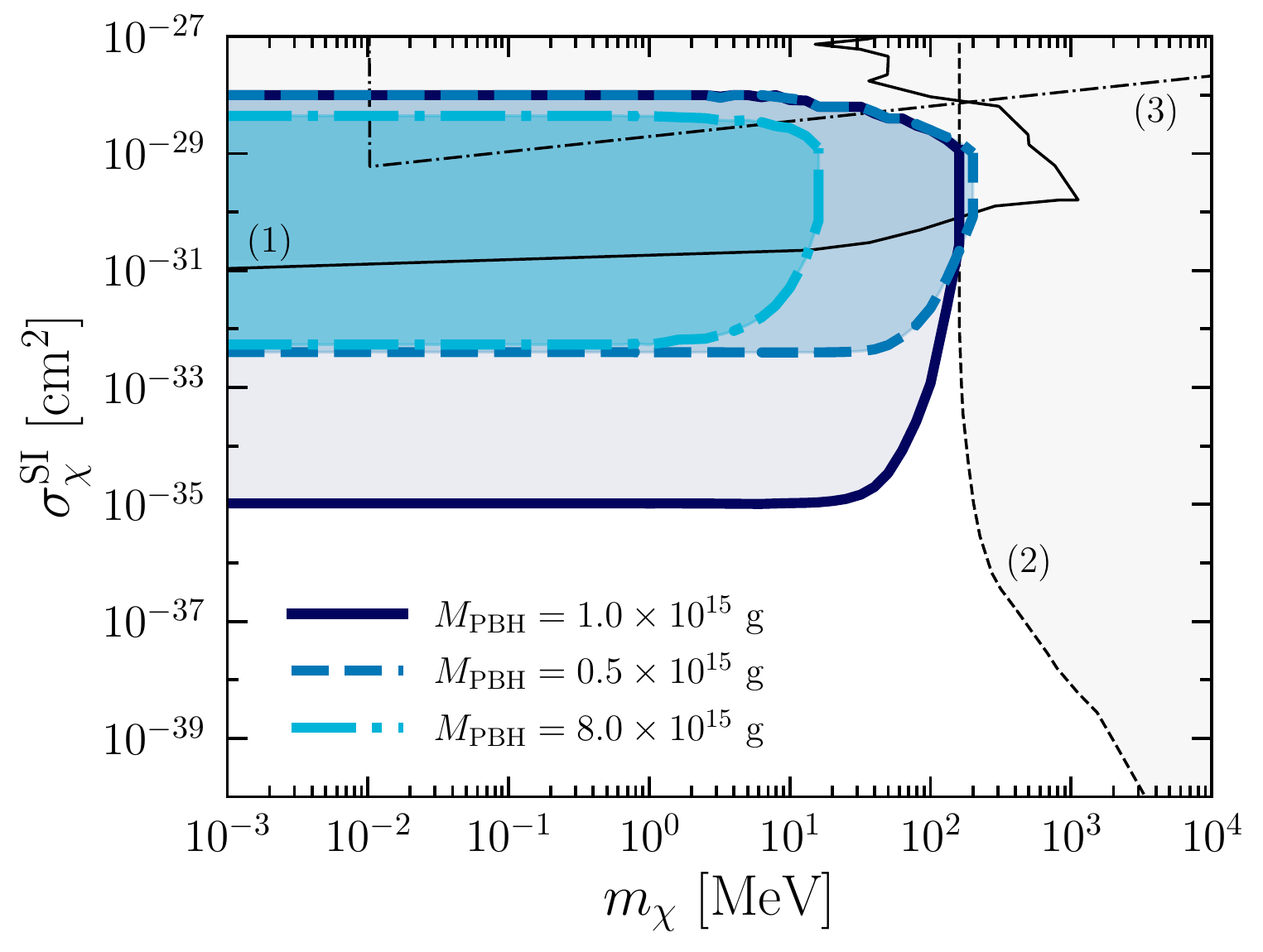}
    \hspace{0.01\textwidth}
    \includegraphics[width=0.485\textwidth]{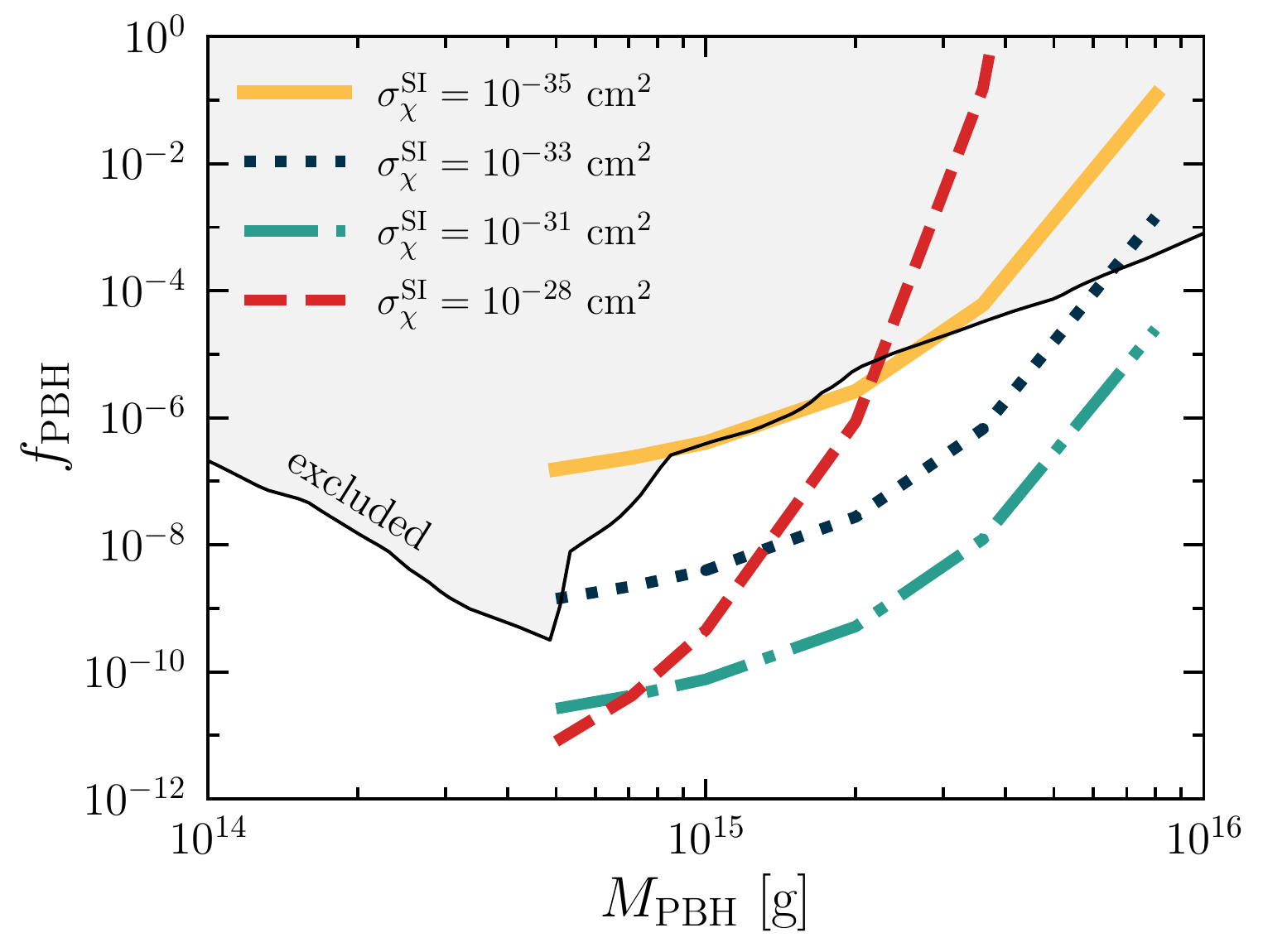}
    \caption{{\bf Constraints on the ePBH-DM parameter space.} We show the regions excluded by XENON1T experiment in the planes $m_\chi$--$\sigma^{\text{SI}}_{\chi}$ (left) and $M_{\text{PBH}}$--$f_{\text{PBH}}$ (right). In the left panel, the excluded regions are obtained for different values of $M_{\text{PBH}}$: for each case, $f_{\text{PBH}}$ is set to the maximum value allowed by present constraints~\cite{Carr:2020gox} (also shown with the thin black line in the right plot). For comparison, we also show previous DM constraints from cosmic-ray boosted DM particles~\cite{Bringmann:2018cvk,Cappiello:2019qsw} (1), from CRESST experiment~\cite{CRESST:2017ues,CRESST:2019jnq} (2), and from cosmology~\cite{Gluscevic:2017ywp,Xu:2018efh,Slatyer:2018aqg,Nadler:2019zrb} (3). In the right panel, the different thick colored lines correspond to the upper bound on $f_\mathrm{PBH}$ obtained for different DM-nucleon cross-section and choosing $m_\chi$ smaller than 1~MeV.}
    \label{fig:XENON}
\end{figure*}

Here we explore for the first time the phenomenological implications of the ePBH-DM scenario in direct detection experiments. Remarkably, we find that even a tiny fraction of evaporating PBHs is enough to give rise to a sizeable flux of boosted light DM particles (see Fig.~\ref{fig:flux}). This translates into a detectable event rate in current experiments such as XENON1T in case DM particles interact with nucleons (see Fig.~\ref{fig:rate}). Hence, the existence of PBHs would imply incredibly strong constraints on the DM parameter space, and viceversa, as shown in Fig.~\ref{fig:XENON}. We find that, assuming PBHs abundances compatible with current bounds, the limits on the spin-independent (SI) DM-nucleon cross-section are improved up to four orders of magnitude. Conversely, the non-observation of the ePBH-DM signal allows us to deduce upper bound on the PBHs abundance a few orders of magnitude stronger than current constraints, depending on the strength of DM-nucleon interactions. \changed{We remark that our constraints do not require the $\chi$ particles to be the dominant DM component.}

{\bf DM flux from evaporating PBHs.--- }
The first step of our work consists in obtaining the DM emission rate from an evaporating PBH. We here make a conservative choice by considering chargeless and spinless PBHs. Spinning PBHs would evaporate faster, thus enhancing the DM emission rate. For the sake of concreteness, we examine the case of DM Dirac fermions with four degrees of freedom, $g_\chi = 4$. However, this framework can be easily extended to scalar and vector DM particles. Differently from Ref.~\cite{Baker:2021btk}, the addition of only one new species to the Standard Model particle content does not significantly alter the standard emission rate of the Hawking radiation. The Hawking temperature of an evaporating PBH with mass $M_\mathrm{PBH}$ is given by~\cite{hawking1975particle, Page:1976df, Page:1977um, PhysRevD.41.3052}
\begin{equation}\label{eq:hawkingtemperature}
    k_\mathrm{B} T_\mathrm{PBH} = 10.6 \left[\frac{10^{15}~{\rm g}}{M_\mathrm{PBH}}\right]~\mathrm{MeV}\,,
\end{equation}
where $k_\mathrm{B}$ is the Boltzmann constant. The differential spectrum per unit time is
\begin{equation}
    \frac{{\rm d}N_\chi}{{\rm d}T{\rm d}t} = \frac{g_\chi}{2\pi}\frac{\Gamma(T, M_\mathrm{PBH})\,}{\exp\left[(T+m_\chi)/k_{\rm B} T_{\rm PBH}\right]+1}\,,
\end{equation}
where $T$ is the kinetic energy of the particle, and $\Gamma$ is the graybody factor, which we compute by means of the \texttt{BlackHawk} code~\cite{Arbey:2019mbc}. 

From the differential spectrum we can compute the flux of DM particles reaching the Earth \changed{for a monochromatic PBH mass distribution}. It consists of a galactic ($\rm gal.$) and an extragalactic ($\rm egal.$) component:
\begin{equation}\label{eq:DMflux}
    \frac{{\rm d}\phi_\chi}{{\rm d}T {\rm d}\Omega}=\frac{{\rm d}\phi^\mathrm{gal.}_\chi}{{\rm d}T {\rm d}\Omega}+\frac{{\rm d}\phi^\mathrm{egal.}_\chi}{{\rm d}T {\rm d}\Omega}\,.
\end{equation}
The first component can be written as
\begin{equation}
    \frac{{\rm d}\phi^\mathrm{gal.}_\chi}{{\rm d}T {\rm d}\Omega}= \frac{f_\mathrm{PBH}}{4\pi\,M_\mathrm{PBH}}\frac{{\rm d} N_\chi}{{\rm d}T{\rm d}t}\int_0^{+\infty}{\rm d}s\,\rho^{\rm NFW}_{\rm DM}(r)\,.
\end{equation}
Here, \changed{the quantity $f_\mathrm{PBH}=\rho_\mathrm{PBH} / \rho_{\rm DM}$ is the fraction of PBHs with respect to the average DM density $\rho_{\rm DM}$ of the Universe}, which is determined by Planck~\cite{Planck:2018vyg}. The galactic flux is proportional to the integral over the line-of-sight distance $s$ of the galactic DM density, \changed{denoted as $\rho^{\rm NFW}_{\rm DM}$}, for which we assume a Navarro-Frank-White profile \cite{Navarro:1996gj}. This is a function of the galactocentric distance $r=(r_{\odot}^2+s^2-2\,s\,r_\odot\cos \ell\cos b)^{1/2}$ with $r_{\odot}=8.5$ kpc and $(b,\ell)$ the galactic coordinates.

For the extragalactic component in Eq.~\eqref{eq:DMflux}, the differential flux takes the expression
\begin{equation}
    \frac{{\rm d}\phi^\mathrm{egal.}_\chi}{{\rm d}T {\rm d}\Omega}=\frac{f_\mathrm{PBH}\,\rho_{\rm DM}}{4\pi\,M_\mathrm{PBH}}\int_{t_\mathrm{min}}^{t_\mathrm{max}}{\rm d}t \,\left[1+z(t)\right]\frac{{\rm d} N_\chi}{{\rm d}T{\rm d}t} \,,
\end{equation}
where we take into account the effect of redshift $z(t)$ on the energy in the DM emission rate, and we integrate from the time of matter-radiation equality ($t_\mathrm{min}$) to the age of the Universe ($t_\mathrm{max}$). Differently from the galactic component, which is enhanced towards the galactic center, the extragalactic DM flux is isotropic.

Given the evaporation temperatures of PBHs from Eq.~\eqref{eq:hawkingtemperature}, DM particles are mainly produced at energies between $1$~MeV and $100$~MeV, for PBHs masses from $10^{14}$ and to $10^{16}$~g. Hence, DM particles lighter than about $1$~MeV are emitted by PBHs with ultra-relativistic velocities. In a sense, the evaporation of PBHs is a new mechanism for boosted DM. In Fig.~\ref{fig:flux}, we show the galactic and extragalactic diffuse flux of DM particles at the Earth, averaged over the whole solid angle, for a reference DM mass $m_\chi=1$~MeV and for different PBH masses. The kinetic energy distribution of particles peaks at different energies depending on the PBH temperature, which can be much larger than the DM mass. For kinetic energies much smaller than the PBH mass, the galactic spectrum flattens since the particles are produced practically at rest. The normalization of the flux is mainly determined by the PBH mass and the concentration of PBHs $f_{\text{PBH}}$. Larger PBH masses lead to slower evaporation rate and lower fluxes. However, at larger PBH masses the constraints on $f_{\text{PBH}}$ are also weaker (see the right plot in Fig.~\ref{fig:XENON}), leading to larger fluxes in Fig.~\ref{fig:flux}. 

{\bf Propagation through Earth and atmosphere.--- }
Direct detection experiments probe DM particles through their potential interactions with nucleons. These processes, however, might also cause an attenuation of the DM flux at the detector position due to the propagation in the Earth and in the atmosphere~\cite{Starkman:1990nj,Mack:2007xj,Kavanagh:2016pyr,Emken:2018run}. For the sake of simplicity, we here model this phenomenon by following the analytical approach described in Ref.~\cite{Bringmann:2018cvk}. In particular, we employ a ballistic-trajectory approximation and compute the energy loss of DM particles reaching the detector. Since the kinetic energy is typically $T \lesssim 100~\mathrm{MeV}$, we can neglect quasi-elastic and inelastic scatterings as well as the nuclear form-factors. Under this assumption, the interaction lengths in the Earth ($\ell^\oplus_\mathrm{int}$) and in the atmosphere ($\ell^{\rm atm}_\mathrm{int}$) read as
\begin{equation}
    \ell^i_\mathrm{int}= \left[\sum_N n^i_\mathrm{N} \sigma_{\chi\mathrm{N}} \frac{2m_\mathrm{N}m_\chi}{(m_\mathrm{N}+m_\chi)^2}\right]^{-1}\,,
\end{equation}
where the sum respectively runs over the most abundant elements in the Earth (from iron to aluminium) and the atmosphere (nitrogen and oxygen) with number density $n^i_\mathrm{N}$, whereas the cross-section is
\begin{equation}
    \sigma_{\chi\mathrm{N}}= \sigma_\chi^{\mathrm{SI}} \mathrm{A}^2 \left(\frac{m_\mathrm{N} (m_\chi+m_\mathrm{p})}{m_\mathrm{p}(m_\chi +m_\mathrm{N})}\right)^2 \,.
\end{equation}
Here we focus on the spin-independent scatterings which feature the standard coherent enhancement with the mass number $\mathrm{A}$. Moreover, we limit ourselves to the case of isospin-singlet structure assuming equal the DM interaction with protons and neutrons. In this framework, DM particles with an initial kinetic energy $T_0$ reach the detector with a smaller kinetic energy $T_d$ after travelling a total geometrical distance $d = d_{\rm atm} + d_\oplus$. We take into account that the XENON1T detector is located in the Gran Sasso National Laboratory at a depth of about $1.4~{\rm km}$~\cite{capuano1998density}. Hence, the actual DM flux reaching the XENON1T detector can be obtained as
\begin{equation}\label{eq:att}
    \frac{{\rm d}\phi_\chi^{d}}{{\rm d}T_{d}{\rm d} \Omega}\approx\frac{4m_\chi^2e^\tau}{\left(2m_\chi + T_d - T_d e^\tau\right)^2}\left(\frac{{\rm d}\phi_\chi}{{\rm d}T {\rm d}\Omega}\bigg|_{T_0}\right)\,,
\end{equation}
where $\tau = d_{\rm atm}/\ell^{\rm atm}_{\rm int} + d_\oplus/\ell^\oplus_{\rm int}$, and the incoming DM flux is evaluated at
\begin{equation}
    T_0\left(T_d\right)=\frac{2m_\chi T_d\,e^\tau}{2m_\chi + T_d - T_d e^\tau}\,.
\end{equation}
For a given incoming direction defined by the galactic coordinates $(b,\ell)$, the distance $d$ depends on the time-dependent longitude of the experiment. Thus, in the following, we consider the average of Eq.~\eqref{eq:att} over a day.

In closing this section, we may stress that the approach of Ref.~\cite{Bringmann:2018cvk} is based on the approximation of collinear, or ballistic, propagation of DM in matter. As pointed out in Ref.~\cite{Cappiello:2019qsw}, this approximation does not apply to the collisions of light DM with heavy nuclei; rather, the propagation of DM happens in a diffusive way, leading to a larger energy loss and a possible reflection of DM from the Earth and the atmosphere. Such a more complicated propagation might lower the ePBH-DM flux detectable at the experiment. In order to understand the impact of the ballistic approximation, we have computed the flux under the most conservative assumption that particles are completely depleted by the scattering (in other words, particles which have undergone even a single scattering are not considered). Even under this extreme assumption, we still find large DM fluxes observable in underground experiments. Therefore, we expect that a more realistic treatment of the DM propagation should not disrupt the scenario presented here. As will be shown later, the ePBH-DM scenario allows us to probe DM-nucleon cross-section at the level of $10^{-35}$~cm$^2$, for which the effects of propagation are marginal, since the atmosphere is completely transparent to DM.

{\bf XENON1T event rate.--- }
We can now compute the event rate expected in the XENON1T detector in the ePBH-DM scenario. In this experiment, the main target for DM scatterings is provided by Xenon nuclei. The differential event rate (number of events per ton year) per Xenon recoil energy $E_r$ can be obtained as
\begin{equation}
    \frac{{\rm d}R}{{\rm d}E_r} = \sigma_{\chi\mathrm{Xe}}\,\mathcal{N}_\mathrm{Xe}\int {\rm d}T_d {\rm d} \Omega\, \frac{{\rm d}\phi_\chi^d}{{\rm d}T_d {\rm d} \Omega} \frac{\Theta(E_r^\mathrm{max}-E_r)}{E_r^\mathrm{max}}\,, 
\end{equation}
where $\mathcal{N}_\mathrm{Xe}$ is the total number of targets, and
\begin{equation}
    E_r^\mathrm{max}=\frac{T_d^2+ 2 m_\chi T_d}{T_d+(m_\chi+m_\mathrm{Xe})^2/(2m_\mathrm{Xe})}
\end{equation}
is the maximum allowed value for the recoil energy, for which we assume a flat distribution. In this estimate, the cross-section $\sigma_{\chi\mathrm{Xe}}$ includes the corresponding nuclear form-factor as provided in Ref.s~\cite{Helm:1956zz,Lewin:1995rx}. Moreover, we consider an exposure of one ton year.

The event rate at XENON1T detector is shown in Fig.~\ref{fig:rate} as a function of the recoil energy. In order to highlight the effect of propagation through the Earth and the atmosphere, we report the ratio of the event rate and the DM-nucleon cross-section $\sigma_\chi^\mathrm{SI}$. In the absence of propagation effects, this ratio would be independent of the cross-section. This can be seen in the case of the solid yellow and dotted blue curve (corresponding to low cross-sections), which are practically identical. For $\sigma^{\text{SI}}_\chi=10^{-31}$~cm$^2$ and $\sigma^{\text{SI}}_\chi=10^{-28}$~cm$^2$, the effects of propagation are more evident, pushing the events to lower recoil energies and weakening the sensitivity to the ePBH-DM signal.

{\bf Results.--- }
In order to probe the ePBH-DM scenario, we need to compare the expected event rate with the current data from XENON1T~\cite{XENON:2018voc}. This experiment has observed no excess of recoil events in the energy window from $E_{r,1} = 4.9~\mathrm{keV}$ to $E_{r,2} = 40.9~\mathrm{keV}$. Following Ref.~\cite{Bringmann:2018cvk}, this can be translated into an upper limit on the total event rate as\changed{
\begin{equation}\label{eq:XENONlim}
    \int_{E_{r,1}}^{E_{r,2}}{\rm d}E_r\frac{{\rm d}R}{{\rm d}E_r} \lesssim \kappa\,\overline{v}_{\rm DM} \,\rho^\odot_{\rm DM}\,\mathcal{N}_{\rm Xe} \left(\frac{\sigma_{\chi {\rm Xe}}}{m_\chi}\right)_{m_\chi \gtrsim 100~{\rm GeV}}\,,
\end{equation}
where $\kappa \simeq 0.23$, $\overline{v}_{\rm DM} \simeq 235$ km/s is the mean DM velocity in the usual scenario, $\rho^\odot_{\rm DM} = 0.3$ GeV/cm$^3$ is the local DM density as assumed by XENON1T Collaboration, and the term in parentheses is $2.5\times 10^{-40} {\rm cm^2/GeV}$ according to the current XENON1T limit for large DM masses~\cite{XENON:2018voc}.}

Equation~\eqref{eq:XENONlim} translates into the limits on the combined parameter space of DM and PBHs shown in Fig.~\ref{fig:XENON}. In particular, the left plot shows the regions of the DM parameter space that are constrained by XENON1T, in the case of three different PBH masses. For each PBH mass, we consider the maximum allowed value for $f_\mathrm{PBH}$ in agreement with current limits~\cite{Carr:2020gox}. The grey region is excluded by previous constraints on sub-GeV DM particles from cosmic-ray upscatterings~\cite{Bringmann:2018cvk,Cappiello:2019qsw}, CRESST experiment~\cite{CRESST:2017ues,CRESST:2019jnq} and cosmology~\cite{Gluscevic:2017ywp,Xu:2018efh,Slatyer:2018aqg,Nadler:2019zrb}. Here, we only focus on the region $m_\chi \geq 10^{-3}~\mathrm{MeV}$ since lighter fermionic DM is highly disfavored. However, our constraints virtually extend to lower DM masses. \changed{For $m_\chi=1~{\rm MeV}$, we have checked that the DM constraints are unchanged for complex scalar DM particles, while they weaken by about $80\%$ for vector DM particles.}

In the right plot, we report the upper bounds on the PBH abundance as a function of the PBH mass, assuming four benchmark values for the DM-nucleon cross-section. Here, we consider DM masses smaller than 1~MeV. For low DM-nucleon cross-sections, the constraints on $f_{\rm PBH}$ are almost inversely proportional to $\sigma_\chi^{\rm SI}$. For $\sigma^\mathrm{SI}_\chi=10^{-28}\mathrm{cm}^{-2}$, instead, the peculiar behaviour of the constraint is due to the considerable effects of the propagation in the Earth and the atmosphere. We emphasize that our constraints are valid for $0.5 \lesssim M_\mathrm{PBH}/(10^{15}~\mathrm{g}) \lesssim 8.0$. Lighter PBHs are almost completely evaporated, whereas heavier ones emit DM particles with too low kinetic energies.

{\bf Discussion.--- }
The ePBH-DM scenario with PBH masses from $10^{14}$ to $10^{16}~{\rm g}$ leads to signatures that are already testable with direct detection experiments. The constraints presented in this work apply to any model with the simultaneous presence of PBHs and DM coupled to nucleons. For low DM-nucleon cross-sections, our results mainly depend on the spectrum of DM from PBHs, and therefore are expected to be robust. For large cross-sections, the different approaches proposed in the literature to describe the DM propagation through Earth and atmosphere might lead to slightly different results. However, this does not influence the possibility of constraining the scenario under consideration.

Since PBHs constitute only a very small fraction of the total energy of the non-relativistic DM in the Universe, the emitted DM particles are expected to not change significantly the properties of the galactic DM distribution and the extragalactic DM density. Indeed, the magnitude of the signal is not connected with a particularly large DM flux from PBHs, but rather with the poorly constrained DM-nucleon cross-sections for sub-GeV DM. Our work shows that cross-sections from $10^{-35}$ to $10^{-31}$~cm$^2$, while allowed by present constraints, may be inconsistent with the existence of PBHs evaporating at present times.

We emphasize that there is a strong degeneracy between the amount of PBHs and the strength of the DM interaction with nucleons. For low cross-sections, where the effects of propagation are negligible, the ePBH-DM signal is indeed proportional to the product $f_{\text{PBH}} \cdot \sigma^{\text{SI}}_{\chi}$. Future observations of evaporating PBHs, for example through the gamma-ray and neutrino burst produced in a PBH evaporation~\cite{Tesic:2012kx,Glicenstein:2013vha,Abdo:2014apa,Archambault:2017asc,Fermi-LAT:2018pfs,HAWC:2019wla,Dave:2019epr,Lopez-Coto:2021lxh}, might confirm their existence and remove this degeneracy, allowing for definite constraints on the DM-nucleon interaction.

\section*{Acknowledgements}
We thank Gennaro Miele and Stefano Morisi for useful comments and discussions. This work was supported by the research grant number 2017W4HA7S ``NAT-NET: Neutrino and Astroparticle Theory Network'' under the program PRIN 2017 funded by the Italian Ministero dell'Universit\`a e della Ricerca (MUR) and by the research project TAsP (Theoretical Astroparticle Physics) funded by the Istituto Nazionale di Fisica Nucleare (INFN).

\bibliography{references}

\end{document}